\documentclass[preprint]{aastex}
\def\simgr{\,\hbox{\hbox{$ > $}\kern -0.8em \lower 1.0ex\hbox{$\sim$}}\,}
\def\simle{\,\hbox{\hbox{$ < $}\kern -0.8em \lower 1.0ex\hbox{$\sim$}}\,}

\shortauthors{THORSTENSEN}
\shorttitle{Anomalously Warm CV Secondary}

\def\asn13cl{ASAS-SN 13cl} 
\def\css1340{CSS J134052.0+151341} 

\begin{document}
\title{\asn13cl : A Newly-Discovered Cataclysmic Binary with an 
Anomalously Warm Secondary  
\footnote{Based on observations obtained at the MDM Observatory, operated by
Dartmouth College, Columbia University, Ohio State University, 
Ohio University, and the University of Michigan.}
}

\author{John R. Thorstensen}
\affil{Department of Physics and Astronomy\\
Dartmouth College\\
Hanover NH, 03755\\
}

\begin{abstract}
The spectrum of the recently discovered cataclysmic variable
star (CV) \asn13cl shows that a secondary star with spectral type 
K4 ($\pm$ 2 subclasses) contributes roughly half the optical light.
The radial velocities of the secondary are modulated on an
orbital period $P_{\rm orb} = 4.86$ hr with a velocity semiamplitude 
$K = 246 \pm 9$ km s$^{-1}$, and the light curve shows 
ellipsoidal variations and an apparent grazing eclipse.  At 
this orbital period, the secondary
stars in most CVs are substantially cooler, with spectral types
near M3.  \asn13cl therefore joins the small group of CVs with 
anomalously warm secondary stars, which apparently form when
the onset of mass transfer occurs after the secondary has undergone
significant nuclear evolution.
\end{abstract}


\section{Introduction}

Cataclysmic variable stars (CVs) are close binaries consisting
of a white dwarf primary that accretes matter from a 
secondary star via Roche lobe overflow.  The secondary is 
more extended than the white dwarf, and usually resembles a
low-mass main-sequence star.  CVs have a rich phenomenology
and have attracted a great deal of observational and 
theoretical interest.  
\citet{warner95} gives a dated but 
useful review of these objects.

Because the secondary fills its Roche lobe, its mean density
is closely constrained by the orbital period, with the secondary's
mean density being larger at shorter periods
\citep{faulkner72}. 
On the main sequence, the mean density rises toward lower masses,
so if the mass-radius relation of the secondary is similar to 
that of main-sequence stars, the orbital period gives a rough proxy 
measurement of the secondary's mass. 
Short period CVs therefore tend to have low-mass secondaries, which
tend to have correspondingly cool temperatures and low
visible luminosities.  \citet{kniggedonor} 
updated the long-known correlation between 
system's orbital period and the secondary's
spectral type.  He found that at a given orbital period, 
secondaries tended to be near the expected
main-sequence surface temperature, or somewhat cooler.

There are exceptions -- a few systems are anomalously
warm, with K-type secondaries in orbits of only a few hours or
less.  These secondaries appear to have undergone 
substantial hydrogen burning during
previous evolution, which increased the
mean molecular weight and hence
changed the mass-luminosity-temperature relation. 
I recently identified an example -- CSS 1340 ,
which has $P_{\rm orb}$ = 2.4 hr; \citet{thorcss13}
details this discovery and 
includes references and further background on these objects.

Here, I report another, less extreme sytem, \asn13cl.
This previously uncatalogued dwarf nova was discovered
in outburst by the ASAS-SN survey
\footnote{The ASAS-SN survey
is described at 
http://www.astronomy.ohio-state.edu/$\approx$assassin/index.shtml} 
on 2013 August 29. 
Table \ref{tab:properties} gives further details on this
object.  

The outburst history of \asn13cl is not well-determined.
It appears consistently, with little obvious variability,
on Digitized Sky Survey images served at Space Telescope Science Institute
\footnote{The many institutions that made this archive possible are
acknowledged at  http://archive.stsci.edu/dss/acknowledging.html.}. 
This area of the sky (at $l = 338, b = -19$ deg) is not covered
by the Catalina Surveys Data Release 2 \citep{drakecrts}.

\section{Observations}

All the observations are from MDM Observatory, on 
Kitt Peak, Arizona.  Spectra were obtained on 2013
Sept.~11, 13, 14, 15, and 17 UT, and photometric 
time series were taken on Sept.~16, 17, and 19 UT.  
The system appeared to have returned to quiescence 
from its recent outburst.

\subsection{Spectroscopy}

The spectra are from the 2.4m Hiltner telescope
and modular spectrograph.  The instrument setup,
observing protocols, and reduction procedures were 
essentially identical to those
described in \citet{thorcss13}.  

\begin{figure}
\plotone{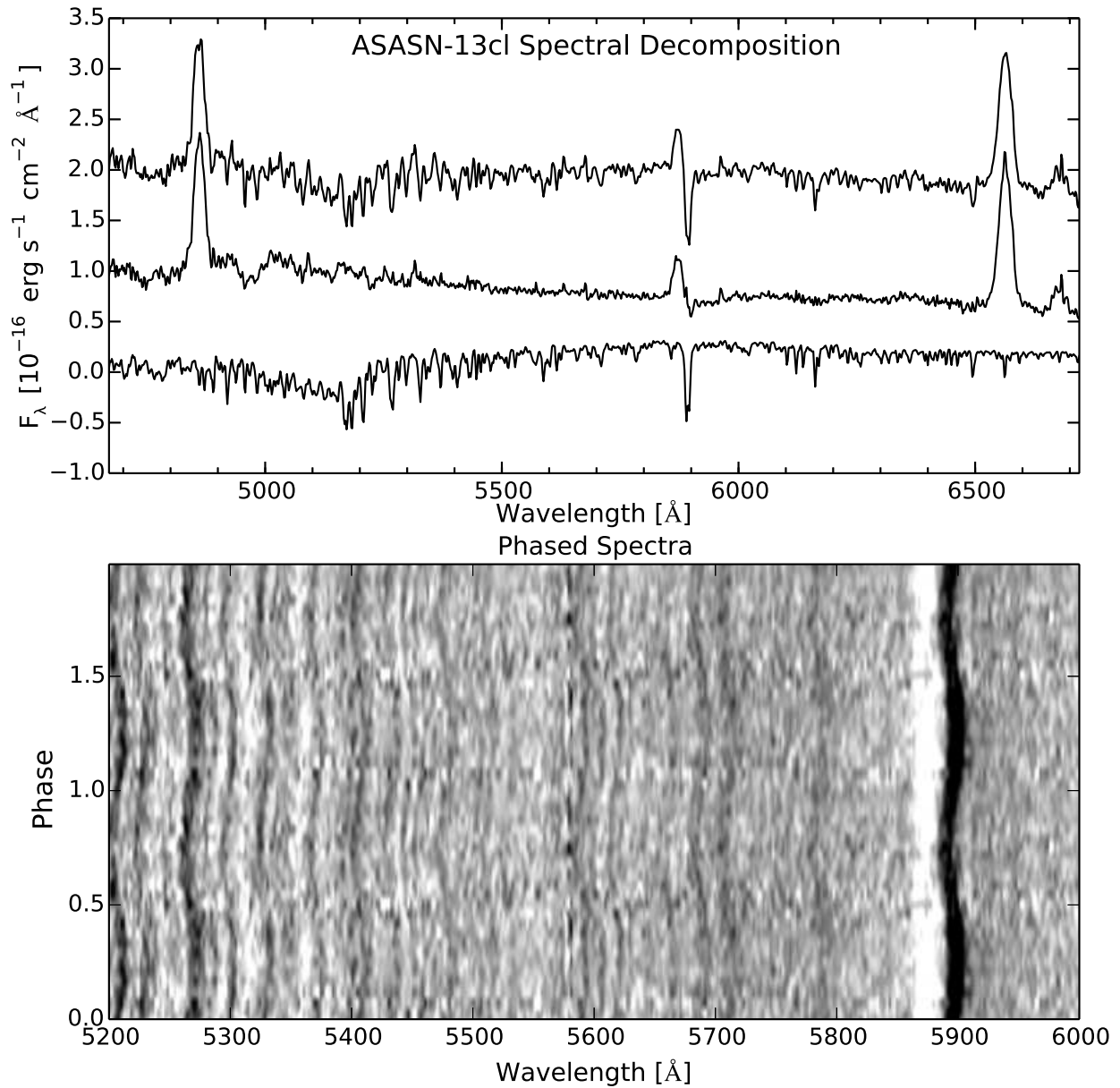}
\caption{{\it Upper panel:} Mean spectrum. The individual spectra were shifted
to zero velocity using the absorption-line ephemeris prior to 
averaging.  The middle trace shows the residual after subtracting
a scaled spectrum of a K4 star Gliese 570a.  The lower trace
shows the scaled K4 spectrum that was subtracted; its zero point is
shifted downward by one unit to avoid overlap.  
{\it Lower panel:} A greyscale representation the portion of the 
spectrum from 5200 to 6000 \AA . 
Each line in the image is derived
from a running average of the rectified spectra that lie near the
nominal orbital phase.  The strong absorption line near 
5900 \AA\ is NaD, and the emission just blueward of this is 
HeI $\lambda 5876$. 
}
\label{fig:specs}
\end{figure}

The mean spectrum (Fig.~\ref{fig:specs}, top panel)
shows the absorption lines of a late-type star together
with the broad emission lines of hydrogen and HeI
commonly seen in dwarf novae at minimum light.
I estimated the
spectral type and light fraction of the secondary
by scaling and subtracting spectra of stars 
classified by \citet{keenan89} and looking
for the best cancellation of late-type features.
This constrained the spectral type to be 
K4 $\pm$ 2 subclasses, with the secondary
contributing roughly half the system's light; the lower
trace in the upper panel of Fig.~\ref{fig:specs} shows
the result of subtracting a scaled K-star from the
observed spectrum.  The SDSS magnitudes in 
Table~\ref{tab:properties} are broadly consistent with this
classification.  The $i - z$ color suggests a type 
near K5 \citep{covey07}, with bluer passbands giving earlier types,
consistent with an increasing contribution from a blue
accretion disk toward shorter wavelengths. 

\begin{figure}
\plotone{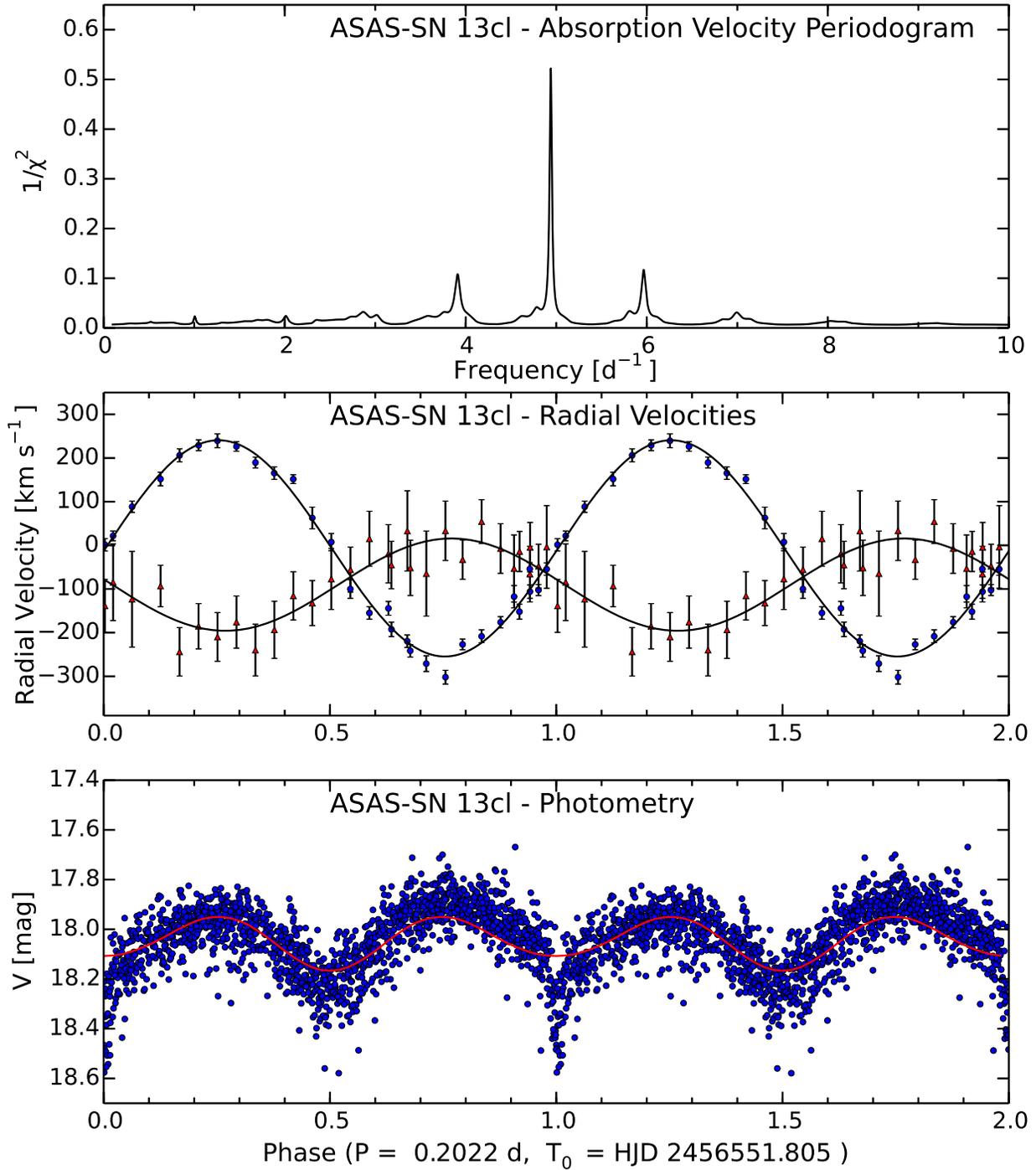}
\caption{{\it Upper panel:} Periodogram of the absorption-line
velocities. 
{\it Middle panel:} Absorption velocities (blue circles) and 
H$\alpha$ (red triangles) emission-line velocities folded on the 
ephemeris given in the text, with the best-fitting sinusoids superposed.  
{\it Lower panel:} V magnitudes, folded on the 
same ephemeris as the middle panel.  The solid curve
is derived from the ellipsoidal variation model described in the
text.
}

\label{fig:folded}
\end{figure}

To measure the (absorption) radial velocity of the K star, I 
cross-correlated the spectra against a velocity-compensated
sum of many late-type spectra, using the 
{\it rvsao} package \citep{kurtzmink}.
The cross-correlation covered from 5100 to 6500 \AA, 
excluding 5800 to 5950 \AA\ to avoid HeI $\lambda$5876
and NaD, which can have an interstellar component.
Table \ref{tab:velocities} gives the velocities.  
A period search of the
resulting timeseries with a `residual-gram' method
\citep{tpst} revealed an unambiguous sinusoidal
modulation at just under 5 cycle d$^{-1}$ 
(Fig.~\ref{fig:folded}).  The H$\alpha$ 
velocities, measured with a convolution routine,
were much noisier than the absorption velocities, but
they do show significant modulation consistent with the
well-defined absorption period.  Table \ref{tab:fits}
gives parameters sinusoidal fits to the velocities.
The velocity modulation of the absorption lines is
clearly visible in 
the lower panel of Fig.~\ref{fig:specs}, which shows a
phase-averaged greyscale representation of the spectra.

 
\subsection{Time-series Photometry}

The time-series photometry is from an Andor Ikon 
DU-937N camera mounted on the MDM 1.3m McGraw-Hill
telescope, as described in \citet{thorcss13}.  Individual
exposures were 30 s, in the $V$ filter.  
In each frame, the program star, a comparison star, and two
check stars were measured; the main comparison star
was 54 arcsec west and 17 arcsec south of the program
star.  Conditions were generally clear but not 
photometric, and exposures in which the comparison 
star's instrumental magnitude more than $\sim 0.4$ mag 
fainter than average were excluded from the analysis.

\begin{figure}
\plotone{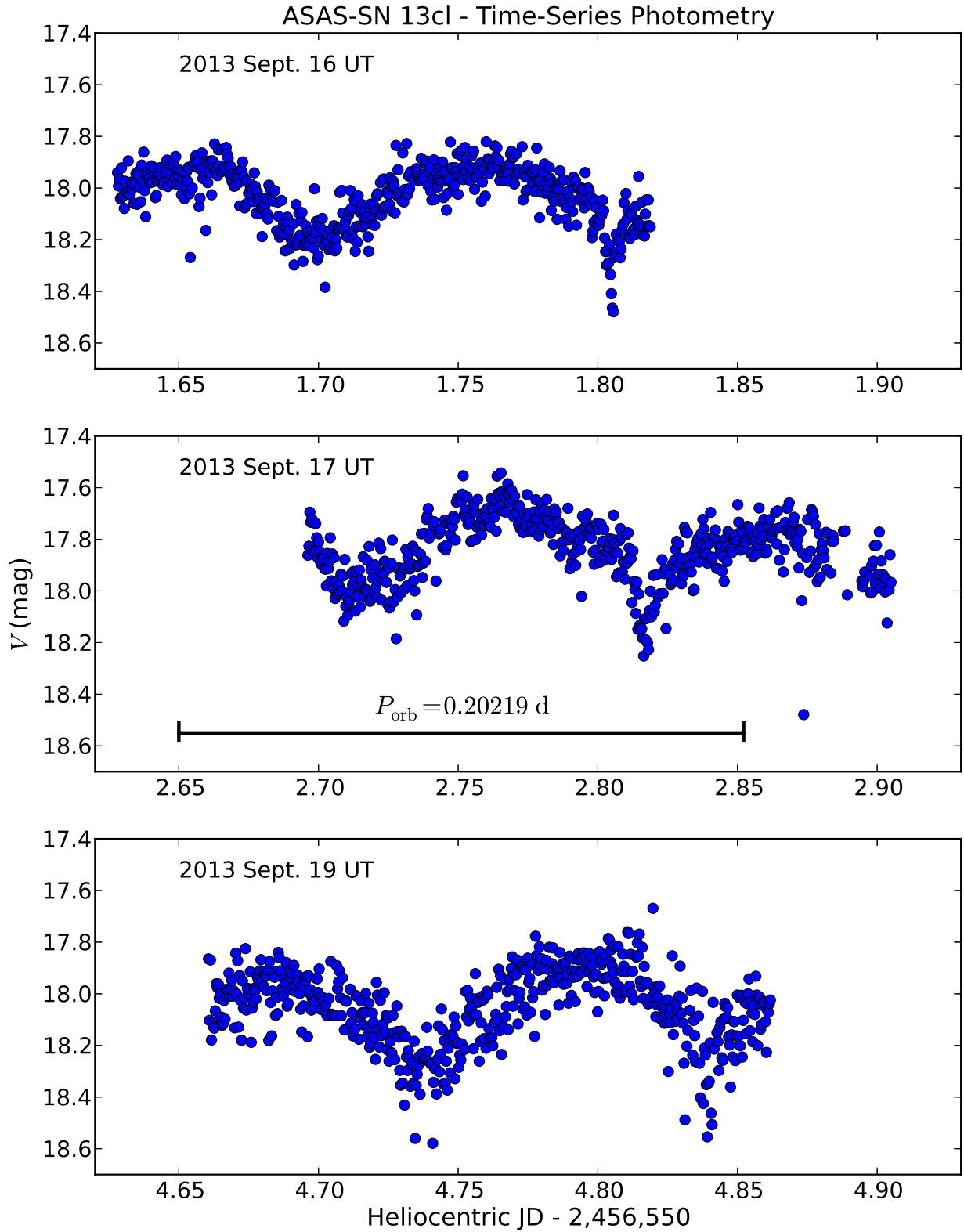}
\caption{Time series photometry of ASAS-SN 13cl from the 
McGraw-Hill 1.3m telescope.  The magnitudes are relative to the 
main comparison star, with the estimated $V$ magnitude of the 
comparison star (15.875) added to set the zero point,
The horizontal bar in the
middle panel shows the length of the orbital period.}
\label{fig:tsphot}
\end{figure}

In the SDSS DR10 (Table~\ref{tab:properties}), 
the program object has $g = 18.265$ and $g-r = +0.683$, while 
the main comparison star has $g = 16.194$ and 
$g-r = +0.545$.  Using transformations derived
by Robert Lupton, these correspond to $V = 17.866$ and
$15.875$ respectively. Fig.~\ref{fig:tsphot} shows
the time series for the three nights, with the
zero-point adjusted to approximate $V$ magnitude, and
the lower panel of Fig.~\ref{fig:folded} shows the 
magnitude plotted against orbital phase.

The photometry shows variation at twice the orbital 
period, consistent with ellipsoidal modulation.
Superposed on this is a sharp dip at inferior
conjunction of the secondary, which is apparently a
grazing eclipse.  The eclipse is clearly 
visible on the first and second nights, but only
just discernible on the third.  Table \ref{tab:eclipses} 
gives the timings of the three eclipses. 

\subsection{Analysis}

{\it Period and Ephemeris.}  The radial velocities
and the grazing eclipses give independent estimates 
of the period, which agree within their uncertainties.  
Adopting the weighted mean period yields
\begin{equation}
\hbox{BJD eclipse} = 2456551.8052(1) + 0.20219(8) E,
\end{equation}
where $E$ is an integer.

{\it Masses.} The grazing eclipse requires an orbital inclination $i$ not
too far from edge-on, and the amplitude of the absorption
line velocity variation $K_{\rm abs}$ should reflect the
secondary star's center-of-mass motion with reasonable
fidelity.  One is less confident of the dynamical interpretation
of the H$\alpha$ velocities (see, e.g., \citealt{marsh87}); 
however, the phase at which 
they cross their mean velocity is $0.515 \pm 0.025$ in the
eclipse ephemeris, consistent with the antiphased motion
expected if they coincide with the white dwarf.  If we take
the velocities as reliable for dynamics, we find 
$q = M_2 / M_1 = K_1 / K_2 = 0.43 \pm 0.08$.  
For $i = 75$ degrees, the mass function then implies
$0.6 {\rm M_{\odot}} < M_{\rm 1} < 0.8 {\rm M_{\odot}}$ 
for the white dwarf, and roughly half that for the 
secondary.  As expected, the secondary is undermassive
for its spectral type; a recent tabulation by Pecaut and
Mamajek\footnote{
available at 
http://www.pas.rochester.edu/$\sim$emamajek/EEM\_dwarf\_UBVIJHK\_colors\_Teff.txt
}
(based on \citealt{pecautandmamajek}) estimates
0.7 M$_\odot$ for typical main-sequence field star at K6, 
the late (hence low-mass) end of our spectral type range.

{\it Distance.} The secondary star's spectral type constrains its surface
brightness, and the Roche lobe geometry combined with the 
orbital period lets us estimate the secondary's radius,
with only a weak dependence on its mass.  The radius and
surface brightness yield an absolute magnitude, which 
together with the observed brightness of the secondary
(and an estimated interstellar extinction)
yields a distance.  All the quantities entering the 
calculation are uncertain, and some are correlated (e.g.,
the light fraction attributed to the secondary decreases
at cooler spectral types, because the metal lines used
in estimating the fraction mostly grow stronger toward
cooler types).  I therefore used a Monte Carlo procedure
to estimate the distance and its uncertainty, 
which explicitly included the correlation between
the secondary's spectral type and its contribution to the
light.  For the K$4 \pm 2$ secondary, I used magnitudes 
centered around $V = 18.5$, and ranging between
extremes of 19.55 (minimum contribution at K6) 
and 18.10 (maximum contribution at K2).  The secondary's
mass (which enters weakly) was taken to be 
uniformly distributed in the range $0.3 \pm 0.15$.  
The $V$-band surface brightness at each spectral
type was computed using a polynomial relation tabulated
by \citet{beuermann06}.  
I assumed $0.02 < E(B-V) < 0.18$, guided by 
the extinction map of \citet{schlegel98}, which gives
a total extinction of $E(B-V) = 0.084$ at this location.  
The Monte Carlo procedure yielded a median distance of
1180 pc, with a 68-percent confidence range from 
990 to 1400 pc.  

{\it Ellipsoidal Variation.} In favorable cases the light
variations of the secondary star can yield useful constraints
on system parameters, especially the inclination.  To exploit this,
\citet{ta05} developed a code to model the light variation of a 
Roche-lobe filling star. \asn13cl, unfortunately, is rather
too faint and messy for fruitful light-curve fitting, but 
nonetheless the lower panel of Fig.~\ref{fig:folded} includes
a representative calculation\footnote{
The curve shown assumes $M_1 = 0.7$ M$_\odot$, 
$M_2 = 0.35$ M$_\odot$, $i = 75$ deg, $T_{\rm eff} = 4620$ K for
the M4 secondary (following \citealt{pecautandmamajek}), and
a contribution from the disk equivalent to an observed 
$V \sim 19.4$, which includes $\sim 0.3$ mag of extinction.
The model is scaled here to a distance of 1100 pc to match the 
observational data; the distance is nearly identical to that 
found earlier by an essentially equivalent method.
} 
which is useful for comparison.  
Several features stand out:
\begin{enumerate}
\item The sharp dip around phase zero departs
significantly from the secondary light curve, indicating
that it is very likely an eclipse.
\item The maxima are unequal, with the second hump
(phase 0.75) higher than the first.  At minimum light
dwarf novae often show a pre-eclipse brightening 
as the `hot spot' where the mass-transfer stream strikes
the disk rotates into view (\citealt{coppejans14} present several
examples of this).  While this has the correct
sense to explain the inequality, the extra light does
not persist until eclipse as it does in other systems, so
it seems more likely that the secondary's leading and 
trailing hemispheres differ in mean brightness.
\item The curve shown has extra light added (assumed
constant through the orbit) to approximate the contributions 
from the rest of the system (accretion disk and so on). 
This tends to flatten the light curve and push the system
to greater computed distances.  The amplitude of the 
computed curve appears to be smaller than that of the data,
suggesting that the extra light may have been overestimated;
however, it is difficult to justify a smaller contribution 
on the basis of the spectral decomposition (Fig.~\ref{fig:specs}). 
\end{enumerate}
Although the data are noisy and the detailed match 
imperfect, the modulation's period and phase confirms
that it is mostly ellipsoidal.

\section{Discussion}

The secondary of \asn13cl is warmer than expected. 
At $P_{\rm orb} = 4.85$ h, the temperature discrepancy 
is not as dramatic as in shorter period systems such as
EI Psc \citet{thoreipsc},
QZ Ser \citep{thorqzser}, 
SDSS J170213.26+322954.1 \citep{littlefair06}, and 
\css1340 \citep{thorcss13}, but it is significant.  

The recent generation of synoptic sky surveys, most notably
the Catalina surveys \citep{drakecrts}, ASAS-SN, and the 
MASTER survey \citep{lipunov2010} have dramatically 
accelerated the rate of discovery of new dwarf novae
(see, e.g., \citealt{breedt14}).  The variability criteria
favor the discovery of large-amplitude outbursts \citep{thorcrts}.
In most CVs with periods less than a few hours, the secondaries
are so faint that the systems can become very faint 
between outbursts, but in systems similar to \asn13cl the
anomalously bright secondary imposes a `floor' on the
total brightness, which can make the amplitude rather modest.
In \asn13cl, the measured amplitude is less than 3 mag, though
it is unlikely to have been caught exactly at maximum.  At 
such relatively small amplitudes, discovery is 
somewhat more difficult.  Also, the known 
examples of warm-secondary dwarf novae have relatively
long inter-outburst times, or have only a single observed
outburst.  For these reasons, it is likely that a substantial
number of these warm-secondary systems await discovery.

\acknowledgments
I gratefully acknowledge support from NSF grant
and AST-1008217.  I also thank the MDM Observatory staff for 
their cheerful and excellent support, and the 
referee for a prompt and useful report.

This paper uses data from the SDSS.
Funding for SDSS-III has been provided by the Alfred P. Sloan Foundation, the
Participating Institutions, the National Science Foundation, and the U.S.
Department of Energy Office of Science. The SDSS-III web site is
http://www.sdss3.org/; further acknowledgments can be found there. 
SDSS-III is managed by the Astrophysical Research Consortium for the
Participating Institutions of the SDSS-III Collaboration.

\begin{deluxetable}{lll}
\tablecolumns{3}
\tablewidth{0pt}
\tablecaption{Properties of \asn13cl}
\tablehead{
\colhead{Property} &
\colhead{Value} &
\colhead{Reference\tablenotemark{a}} \\
}
\startdata
$\alpha_{2000}$ & 21$^{\rm h}$ 38$^{\rm m}$ 05$^{\rm s}$.046  & SDSS \\
$\delta_{2000}$ & +26$^\circ 38' 19".71$ & \\[1.2ex]
Outburst date & 2013-08-29.35 & ASAS-SN \\
Outburst magnitude & 15.66 & \\[1.2ex]
u & $19.097 \pm 0.028$ & SDSS \\  
g & $18.265 \pm 0.008$ & \\
r & $17.582 \pm 0.016$ & \\
i & $17.287 \pm 0.007$ & \\
z & $17.055 \pm 0.014$ & \\[1.2ex]
V & 17.87 \tablenotemark{b}  &  \\[1.2ex]
J & $15.137 \pm 0.053$ &  2MASS \\
H & $14.814 \pm 0.063$ &  \\
K & $14.701 \pm 0.079$ & \\[1.2ex]
\enddata
\tablenotetext{a}{References are as follows: SDSS is the 
Sloan Digital Sky Survey Data Release 10; 
2MASS is described by
\citet{2mass}. 
} 
\tablenotetext{b}{The $V$ magnitude is computed using 
$$V = g - 0.5784 (g - r) - 0.0038,$$
an approximation derived by R. H. Lupton and cited at
{\tt http://www.sdss.org/dr7/algorithms/sdssUBVRITransform.html}
}
\label{tab:properties}
\end{deluxetable}

\begin{deluxetable}{lrrrr}
\tablecolumns{5}
\tablewidth{0pt}
\tablecaption{Radial Velocities}
\tablehead{
\colhead{Time} &
\colhead{$v$ (absn)} &
\colhead{$\sigma$} & 
\colhead{$v$ (emn)} &
\colhead{$\sigma$} \\
\colhead{} &
\colhead{[km s$^{-1}$]} &
\colhead{[km s$^{-1}$]} &
\colhead{[km s$^{-1}$]} &
\colhead{[km s$^{-1}$]} \\
}
\startdata
56546.9338  & $ -118$ & $  26$ & $  -53$ & $  77$ \\
56546.9408  & $  -54$ & $  15$ & $  -66$ & $  64$ \\
56548.9009  & $ -192$ & $  17$ & $  -45$ & $  54$ \\
56548.9093  & $ -242$ & $  14$ & $  -52$ & $  63$ \\
56549.7790  & $  -54$ & $  14$ & $   -4$ & $  95$ \\
56549.7875  & $   22$ & $  11$ & $  -85$ & $  88$ \\
56549.7960  & $   88$ & $  13$ & $ -123$ & $ 110$ \\
56549.8087  & $  152$ & $  16$ & $  -93$ & $  48$ \\
56549.8172  & $  206$ & $  15$ & $ -244$ & $  55$ \\
56549.8257  & $  229$ & $  12$ & $ -185$ & $  52$ \\
56549.8341  & $  239$ & $  16$ & $ -210$ & $  56$ \\
56549.8426  & $  226$ & $  11$ & $ -176$ & $  60$ \\
56549.8511  & $  190$ & $  12$ & $ -240$ & $  59$ \\
56549.8595  & $  165$ & $  14$ & $ -194$ & $  65$ \\
56549.8680  & $  152$ & $  10$ & $ -116$ & $  55$ \\
56549.8765  & $   63$ & $  25$ & $ -132$ & $  52$ \\
56549.8850  & $    7$ & $  20$ & $  -77$ & $  70$ \\
56549.8935  & $ -100$ & $  21$ & $  -56$ & $  52$ \\
56549.9020  & $ -155$ & $  14$ & $   15$ & $  63$ \\
56549.9105  & $ -144$ & $  16$ & $  -19$ & $  67$ \\
56549.9189  & $ -220$ & $  14$ & $   33$ & $  92$ \\
56549.9274  & $ -271$ & $  18$ & $  -65$ & $  97$ \\
56549.9360  & $ -302$ & $  16$ & $   33$ & $  68$ \\
56550.7524  & $ -227$ & $  12$ & $  -33$ & $  45$ \\
56550.7609  & $ -208$ & $  15$ & $   54$ & $  50$ \\
56550.7694  & $ -176$ & $  13$ & $   -7$ & $  57$ \\
56550.7778  & $ -152$ & $  17$ & $  -14$ & $  46$ \\
56550.7863  & $ -102$ & $  13$ & $  -48$ & $  50$ \\
56550.7948  & $    1$ & $  14$ & $ -139$ & $  61$ \\
56552.8044  & $ -106$ & $  15$ & $   -4$ & $  57$ \\
\enddata
\tablecomments{ Times given as barycentric
Julian dates of mid-integration, minus 2 400 000.
The time base is UTC.
}
\label{tab:velocities}
\end{deluxetable}

\begin{deluxetable}{lrrrrrr}
\tablecolumns{7}
\tablewidth{0pt}
\tablecaption{Fits to Radial Velocities}
\tablehead{
\colhead{Description} &
\colhead{$T_0$} &
\colhead{$P$} & 
\colhead{$K$} & 
\colhead{$\gamma$} & 
\colhead{$N$} & 
\colhead{$\sigma$} \\ 
\colhead{} &
\colhead{(BJD)} &
\colhead{(d)} & 
\colhead{(km s$^{-1}$)} & 
\colhead{(km s$^{-1}$)} & 
\colhead{} & 
\colhead{(km s$^{-1}$)} \\
}
\startdata
Absorption & 56549.7839(11) & 0.20246(20) &  246(9) & $-6(6)$ & 30 &  20 \\ 
           & 56549.7841(12) & [0.20219] &  248(9) & $-7(6)$ & 30 &  22 \\[1.2ex]
H$\alpha$ emission & 56549.888(5) & 0.2026(9) &  105(17) & $-91(12)$ & 30 &  40 \\
           & 56549.888(5) & [0.20219] &  106(17) & $-90(12)$ & 30 &  40 \\
\enddata
\tablecomments{Parameters of best-fit sinusoids of the 
form $v(t) = \gamma + K \sin[2 \pi (t - T_0) / P].$  The
epoch $T_0$ is expressed as a barycentric Julian day.  In the 
second line for each fit, the period is fixed at the weighted
mean derived from the radial velocities and eclipse timings. 
}
\label{tab:fits}
\end{deluxetable}

\begin{deluxetable}{lll}
\tablecolumns{3}
\tablewidth{0pt}
\tablecaption{Eclipse Timings}
\tablehead{
\colhead{$E$} &
\colhead{Barycentric JD} &
\colhead{$\sigma$} \\
} 
\startdata
0 & 2456551.8052 & 0.0003 \\ 
5 & 2456552.8164 & 0.0007 \\
15 & 2456554.8367 & 0.0012 \\
\enddata
\tablecomments{$E$ is an integer cycle 
count.  Times given are barycentric
Julian dates of mid-eclipse. 
The time base is UTC.  The last column 
is the estimated one-sigma uncertainty of
the timing, in days.
}
\label{tab:eclipses}
\end{deluxetable}

%

\end{document}